\documentclass{article}
\input epsf

\newcommand{\ints}{\int_0^{\hspace{.5ex}\infty}
{^{\hspace{-2.8ex}\textstyle '}}\hspace{2ex}}
\newcommand{\intss}{\int_0^{\hspace{1ex}\infty}
{^{\hspace{-3.5ex}\textstyle ''}}\hspace{2ex}}
\newcommand{\Ints}{\int_0^{\hspace{.5ex}\infty}
{^{^{^{\hspace{-2.8ex}\textstyle '}}\hspace{1.5ex}}}}

\newcommand{\sumprime}{\hspace{-.5ex}{^{\textstyle'}}}
\newcommand{\bbox}[1]{\mbox{\boldmath $#1$}}

\begin{document}

\renewcommand{\thefootnote}{\fnsymbol{footnote}}
\begin{flushleft}
\vspace* {1.2 cm}
{\Large\bf
{
Dipole-Dipole Coupling in the Presence of Dispersing and
Absorbing Bodies
}$\!\!$\footnote{Proceedings of 9th
Central-European Workshop on Quantum Optics, Szeged, 2002,
to appear in Fortschritte der Physik}
}\\
\vskip 1truecm
{\large\bf
{
D.-G. Welsch$^1$), Ho Trung Dung$^{1,2}$) and L. Kn\"{o}ll$^1$)
}
}\\
\vskip 5truemm
{
$^1$) Theoretisch-Physikalisches Institut,
Friedrich-Schiller-Universit\"{a}t Jena,\\
Max-Wien-Platz 1, 07743 Jena, Germany\\
$^2$) Institute of Physics, National Center for Natural Sciences and
Technology,\\
1 Mac Dinh Chi Street, District 1, Ho Chi Minh City, Vietnam
}

\end{flushleft}
\vskip 0.5truecm
{\bf Abstract:\\}

{
\noindent
An effective Hamiltonian and equations of motion for treating
both the resonant dipole-dipole interaction between two-level
atoms and the resonant atom-field interaction are derived, which can
suitably be used for studying the influence of arbitrary dispersing
and absorbing material surroundings on these interactions.
It is shown that the dipole-dipole interaction
and the atom-field interaction, respectively, are closely related to
the real part and the imaginary part of the
(classical) Green tensor. The theory is applied to the study of the
transient behavior of two atoms that initially share a single
excitation, with special emphasis on the role of the two
competing processes of virtual and real photon exchange in the
energy transfer between the atoms. To illustrate the powerfulness
of the theory, specific results for the case of the atoms being
near a dispersing and absorbing microsphere are presented.
In particular, the regimes of weak and strong atom-field coupling
are addressed.

}
\vskip 0.1 cm
\noindent
PACS: 42.50.Ct, 42.50.Fx, 42.60.Da, 82.20.Rp

\section{Introduction}
\label{sec1}

The progress in trapping single atoms in microresonator-type
equipments has offered novel possibilities of studying
fundamental quantum processes on the level of a very
few atoms. Let us consider two two-level atoms and assume
that one of them is initially prepared in the upper state.
Various processes are possible. (i) The excited atom can
spontaneously emit a real photon that leaves
the resonator or is reabsorbed by the atom. (ii) The emitted
photon is lost because of material absorption. (iii) The excited
atom undergoes radiationless decay because of material absorption.
(iv) The emitted photon is absorbed by the second atom that is
initially in the lower state, i.e., the excitation energy is exchanged
between the two atoms. (v) The excitation energy is exchanged
between the atoms via virtual photon emission and absorption,
which is commonly termed resonant dipole-dipole interaction.
All these processes sensitively depend on the
electromagnetic-field structure which is controlled by the
material surroundings forming the microresonator.

In the regime of weak atom-field coupling, the mutual interaction
of atoms has typically been described by an effective two-body
potential involving atomic variables only. By analyzing a
one-dimensional cavity model, it has been shown that this
concept may fail to give a correct description of the interaction
at least in the strong-coupling regime, where the electromagnetic
field degrees of freedom can no longer be eliminated \cite{Goldstein97}.
In order to describe both the resonant dipole-dipole interaction
and the resonant atom-field interaction of two
two-level atoms in a high-$Q$ cavity, an effective
Hamiltonian has been proposed \cite{Kurizki96}.
Apart from the fact that the introduced coupling parameters
are not specified and their relation to each other thus
remains unclear (also see \cite{Kurizki88,Zheng96}),
material absorption cannot be taken into account, because
of the underlying concept of mode decomposition.

\section{Hamiltonian}
\label{sec2}

In order to treat the problem more rigorously, let us start from
the multipolar-coupling Hamiltonian for $N$ two-level atoms
[positions ${\bf r}_A$, transition frequencies $\omega_A$,
transition dipole moments ${\bf d}_A$ ($A$ $\!=$ $\!1,2,...,N$)] that
interact with the electromagnetic field via electric-dipole transitions
in the presence of dispersing and absorbing bodies \cite{Knoll01,Ho01b}:
\begin{equation}
\label{e2}
   \hat{H} = \int \!{\rm d}^3{\bf r}
   \! \int_0^\infty \!\!{\rm d}\omega \,\hbar\omega
   \,\hat{\bf f}^\dagger({\bf r},\omega){}\hat{\bf f}({\bf r},\omega)
    + \sum_{A} {\textstyle{1\over 2}}\hbar\omega_A \hat{\sigma}_{Az}
    - \sum_{A} \int_0^\infty \!\!{\rm d}\omega \left[
    \hat{\bf d}_A
    \underline{\hat{\bf E}}({\bf r}_A,\omega)
    + {\rm H.c.}\right]
\end{equation}
($\hat{\bf d}_A$ $\!=$ $\!{\bf d}_A\hat{\sigma}_A$ $\!+$
$\!{\bf d}_A^\ast \hat{\sigma}_A^\dagger$), where
\begin{equation}
\label{e3}
     \underline{\hat{\bf E}}({\bf r},\omega)
     = i \sqrt{\frac{\hbar}{\pi\varepsilon_0}}
     \frac{\omega^2}{c^2}
     \int {\rm d}^3{\bf r}'
     \sqrt{\varepsilon_{\rm I}({\bf r}',\omega)}
     \,\bbox{G}({\bf r},{\bf r}',\omega)
     {}\hat{\bf f}({\bf r}',\omega).
\end{equation}
Here, $\bbox{G}({\bf r},{\bf r}',\omega)$ is the classical
Green tensor satisfying the inhomogeneous Helmholtz equation
\begin{equation}
\label{e3a}
\left[\frac{\omega^2}{c^2}\,\varepsilon({\bf r},\omega)
-\mbox{\boldmath $\nabla$}\times\mbox{\boldmath $\nabla$}
\times\right]\mbox{\boldmath $G$}({\bf r},{\bf r}',\omega)
= - \delta({\bf r}-{\bf r}'),
\end{equation}
with
$\varepsilon({\bf r},\omega)$
$\!=$ $\!\varepsilon_{\rm R}({\bf r},\omega)$
$\!+$ $\!i\varepsilon_{\rm I}({\bf r},\omega)$ 
being the complex (Kramers-Kronig consistent)
permittivity of the material surroundings.
The fields $\hat{\bf f}({\bf r},\omega)$ and
$\hat{\bf f}^\dagger({\bf r},\omega)$ are bosonic ones which
play the role of the fundamental variables of the electromagnetic
field and the medium, including a reservoir necessarily associated
with the losses in the medium.
It is not difficult to see that $\hat{\bf f}({\bf r},\omega)$
obeys the Heisenberg equation of motion
\begin{equation}
\label{e3.1}
       \dot{\hat{\bf f}}({\bf r},\omega)
       = -i\omega \hat{\bf f}({\bf r},\omega)
       +\, {\omega^2\over c^2}
       \sqrt{\varepsilon_{\rm I}({\bf r},\omega)
       \over  \hbar\pi\varepsilon_0}
       \sum_{A} \hat{\bf d}_{A}
       \bbox{G}^\ast({\bf r}_{A},{\bf r},\omega).
\end{equation}

To treat the off-resonant atom-field interaction, we proceed as
follows. (i) The $\omega$-integrals in Eq.~(\ref{e2}) are decomposed
into on-resonant parts (denoted by $\ints {\rm d\omega}\ldots$) and
off-resonant parts (denoted by$\intss {\rm d\omega}\ldots$).
(ii) For $\hat{\bf f}(\omega)$ [and $\hat{\bf f}^\dagger(\omega)$]
in the off-resonant part of the third term on the right-hand side
in Eq.~(\ref{e2}), the formal solution of Eq.~(\ref{e3.1})
is substituted, i.e.,  
\begin{equation}
       \hat{\bf f}({\bf r},\omega,t)
       = \hat{\bf f}_{\rm free}({\bf r},\omega,t)
       + {\omega^2\over c^2}
        \sqrt{\varepsilon_{\rm I}({\bf r},\omega) \over  \hbar\pi\varepsilon_0}
        \sum_{A} \int_0^t {\rm d} t'\,
        \hat{\bf d}_{A}(t')\,\bbox{G}^\ast({\bf r}_{A},{\bf r},\omega)
        e^{-i\omega(t-t')},
\end{equation}
where $\hat{\bf f}_{\rm free}({\bf r},\omega,t)$ evolves freely.
(iii) In the resulting expression, slowly varying atomic
operators are put in front of the $t'$-integrals, so that the
$t'$-integrals can be performed to (approximately) yield
$\zeta$-functions which are (approximately) replaced
by their principal-value parts (because of the off-resonance
condition). (iv) The expectation value
with respect to the off-resonant medium-assisted field is
taken, by assuming that it is in the vacuum state.

If the differences between the atomic transition frequencies are small
compared to the frequency scale of variation of the Green tensor,
then the procedure outlined changes the Hamiltonian (\ref{e2}) into
the effective Hamiltonian
\begin{eqnarray}
\label{e11}
\lefteqn{
   \hat{H}_{\rm eff} = \int \!{\rm d}^3{\bf r}
   \Ints {\rm d}\omega \,\hbar\omega
   \,\hat{\bf f}^\dagger({\bf r},\omega){}\hat{\bf f}({\bf r},\omega)
    + \sum_{A} {\textstyle{1\over 2}}\hbar\tilde{\omega}_A \hat{\sigma}_{Az}
}
\nonumber\\&&\hspace{2ex}
    - \sum_{A,A'}\sumprime 
    \hbar \delta_{A^\ast A'}
       \hat{\sigma}^\dagger_A\hat{\sigma}_{A'}
    - \sum_{A} \Ints {\rm d}\omega \left[
    \hat{\bf d}_A  \underline{\hat{\bf E}}({\bf r}_A,\omega)
    + {\rm H.c.}\right],
\end{eqnarray}
where the notation $\sum_{A,A'}'$ indicates that
\mbox{$A$ $\!\neq$ $A'$}. The $\tilde{\omega}_A$
are the shifted transition frequencies,
and the $\delta_{A^\ast A'}$ ($A$ $\!\neq$ $A'$) are the
resonant dipole-dipole coupling strengths,
\begin{equation}
\label{e10}
    \tilde{\omega}_A = \omega_A-\delta_{A^\ast A},
\end{equation}
\begin{equation}
\label{e3.8}
       \delta_{A^\ast A} = \delta^-_{A^\ast A} -  \delta^+_{A^\ast A} ,
\end{equation}
\begin{equation}
\label{e8}
       \delta_{A^\ast A'}
       = \delta^-_{A^\ast A'} +  \delta^+_{A^\ast A'},
\end{equation}
\begin{equation}
\label{e3.7}
       \delta^{-(+)}_{AA'} 
       = {{\cal P}\over \pi\hbar\varepsilon_0}
       \int_0^\infty {\rm d}\omega \,{\omega^2\over c^2}
       { {\bf d}_A \,{\rm Im}\,\bbox{G} ({\bf r}_A,{\bf r}_{A'},\omega)
       \,{\bf d}_{A'}
       \over \omega-(+)\tilde{\omega}_{A'}}
\end{equation}
[${\cal P}$ -- principal value]. The notation $A^\ast$ ($A'^\ast$)
means that ${\bf d}_A$ (${\bf d}_{A'}$) in
Eq.~(\ref{e3.7}) has to be replaced with its complex conjugate
${\bf d}_A^\ast$ (${\bf d}_{A'}^\ast$).
Recalling the Kramers-Kronig relation for the Green tensor,
we may approximately rewrite Eq.~(\ref{e8}) as
\begin{equation}
\label{e10.2}
       \delta_{A^\ast A'} 
       = {\tilde{\omega}_{A'}^2\over \hbar\varepsilon_0 c^2}\,
       {\bf d}_A^\ast {\rm Re}\,\bbox{G}({\bf r}_A,{\bf r}_{A'},
       \tilde{\omega}_{A'}){\bf d}_{A'} ,
\end{equation}
which reveals that the resonant dipole-dipole interaction is
closely related to the real part of the Green tensor.
Accordingly, from Eqs.~(\ref{e3.7})
and (\ref{e3.8}) it (approximately) follows that
\begin{equation}
\label{e10.1}
       \delta_{A^\ast A} =
       {\omega_A^2\over \hbar\varepsilon_0 c^2}\,
       {\bf d}_A^\ast {\rm Re}\,\bbox{G}({\bf r}_A,{\bf r}_A,
       \tilde{\omega}_{A}){\bf d}_A
       -  2\delta^+_{A^\ast A}.
\end{equation}
Note that the Green tensor can typically written as a sum
of the vacuum Green tensor $\bbox{G}_{\rm V}$ and a reflection
part $\bbox{G}_{\rm R}$. Due to the singularity of
${\rm Re}\,\bbox{G}_{\rm V}$ at equal space points,
Eq.~(\ref{e10.1}) actually applies to the reflection part only.
The vacuum part can be thought of as being already included
in $\omega_A$.

\section{Equations of motion}
\label{sec3}

Let us assume that the atoms
initially share a single excitation while the
medium-assisted electromagnetic field is in the vacuum state.
In the Schr\"{o}dinger picture, we then may write, on omitting
off-resonant terms, the state vector of the system in the form of
\begin{equation}
\label{e4.1}
    |\psi(t)\rangle = \sum_A C_A(t)
    e^{-i(\tilde{\omega}_A-\bar{\omega})t}
    |U_A\rangle |\{0\}\rangle
     + \int {\rm d}^3{\bf r} \Ints \!{\rm d}\omega\,
     {\bf C}_{L}({\bf r},\omega,t)
     e^{-i (\omega-\bar{\omega})t}
     |L\rangle\hat{\bf f}^\dagger({\bf r},\omega)|\{0\}\rangle
\end{equation}
($\bar{\omega}$ $\!=$ $\!$ $\!\frac{1}{2}\sum_A\tilde{\omega}_A$).
Here, $|U_A\rangle$ is the atomic state with the $A$th atom
in the upper state and all the other atoms in the lower state,
and $|L\rangle$ is the atomic state with all atoms in the lower state.
Accordingly, $|\{0\}\rangle$ is the vacuum state of the
rest of the system, and
$\hat{f}^\dagger_i({\bf r},\omega) |\{0\}\rangle$
is the state, where a single quantum is excited.

Basing on the Hamiltonian (\ref{e11}),
it is straightforward to derive the equations of motion
for the slowly varying probability amplitudes
$C_A(t)$ and ${\bf C}_{L}(t)$.
By formally integrating the equation for ${\bf C}_{L}(t)$
under the initial condition that \mbox{${\bf C}_{L}(t=0)$ $\!=$ $\!0$},
and substituting the formal solution into the equation for
$C_A(t)$, we obtain the following
system of coupled integrodifferential equations for the $C_A(t)$:
\begin{equation}
\label{e9}
        \dot{C}_A(t) =
	  \sum_{\scriptstyle A' \atop \scriptstyle A'\neq A}
	i\delta_{A^\ast A'}
        e^{i(\tilde{\omega}_A-\tilde{\omega}_{A'})t}
        C_{A'}(t)
        + \sum_{A'}
        \int_0^t {\rm d}t'
        \Ints \!{\rm d}\omega
         K_{A^\ast A'}(t,t';\omega)
        \, C_{A'}(t'),
\end{equation}
where
\begin{equation}
\label{e6}
        K_{AA'}(t,t';\omega)
        = -\frac{1} {\hbar\pi\varepsilon_0}
        \biggl[ {\omega^2\over c^2}
        e^{-i(\omega-\tilde{\omega}_A)t}
        e^{i(\omega-\tilde{\omega}_{A'})t'}
       {\bf d}_A {\rm Im}\,\bbox{G}({\bf r}_A,{\bf r}_{A'},\omega)
       {\bf d}_{A'} \biggr].
\end{equation}
Equations (\ref{e9}) and (\ref{e6}) reveal that the resonant
atom-field interaction is closely related to the imaginary part
of the Green tensor.
Since the real part and the imaginary part of the Green tensor
are related to each other, it is clear that the resonant dipole-dipole
coupling strengths and the resonant atom-field coupling strengths
cannot be chosen independently from each other in general.

\section{Two atoms near a microsphere}
\label{sec4}

\begin{figure}[t!]
\begin{center}
\epsfxsize=1.0\textwidth
\epsfbox{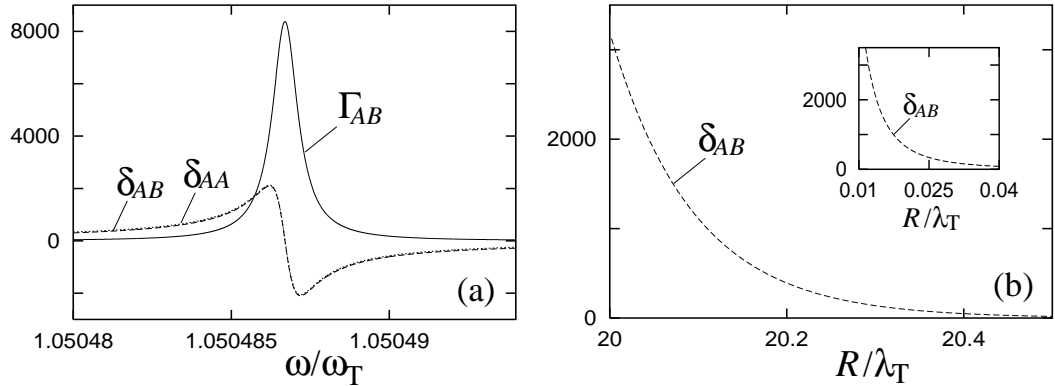}
\caption{
$\Gamma_{AB}$, $\delta_{AB}$ (dashed line),
and $\delta_{AA}$ (dotted line), which almost
coincides with $\delta_{AB}$, in units of
the free-space spontaneous decay rate $\Gamma_0$,
are shown for two atoms situated near
a dielectric microsphere of single-resonance Drude-Lorentz-type
[$\omega_{\rm T}$, transverse frequency;
\mbox{$\omega_{\rm P}$ $\!=$ $\!0.5\,\omega_{\rm T}$},
plasma frequency;
$\gamma$ $\!=$ $\!10^{-6}\,\omega_{\rm T}$, absorption parameter;
$d$ $\!=$ $\!20\lambda_{\rm T}$, sphere diameter ($\lambda_{\rm T}$
$\!=$ $\!2\pi c/\omega_{\rm T}$);
$\Delta r_A$ $\!\equiv$ $\!r_A$ $\!-$ $\!0.5\,d$
$\!=$ $\!\Delta r_B$ $\!\ge$ $\!10^{-3}\,\lambda_{\rm T}$,
distance of the atoms from the sphere surface;
${\bf d}_A$ $\!=$ $\!{\bf d}_B$, radially oriented
real transition dipole moments;
(a) $\Delta r_A$ $\!=$ $\!0.02\,\lambda_{\rm T}$;
\mbox{$R$ $\!=$ $\!20.04\,\lambda_{\rm T}$}, interatomic distance;
(b) $\omega$ $\!=$ $\!1.05048621\,\omega_{\rm T}$].
For comparison, the inset in (b) shows the free space case.
\label{f1}
}
\end{center}
\end{figure}
%
\begin{figure}[!ht]
\begin{center}
\epsfxsize=.5\textwidth
\epsfbox{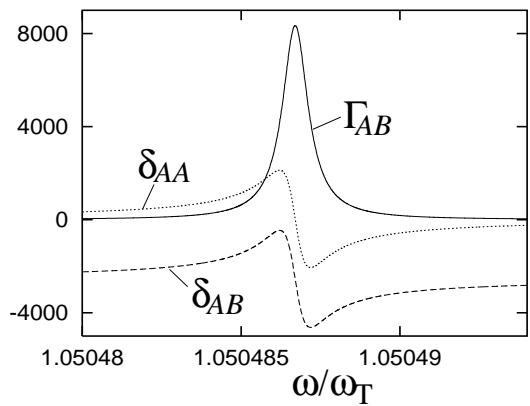}
\caption{
Same as in Fig.~\ref{f1}(a), but for a short interatomic distance
$R$ $\!=$ $\!0.01\,\lambda_{\rm T}$.
\label{f2}
}
\end{center}
\end{figure}
%
To illustrate the theory, let us consider two identical
atoms ($A$ and $B$) near a dispersing and absorbing microsphere
and assume that they have equivalent positions and dipole
orientations with respect to the sphere such that the relations
${\cal K}_{A^\ast A}={\cal K}_{B^\ast B}$ and
${\cal K}_{A^\ast B}={\cal K}_{B^\ast A}$ are valid, where
\begin{equation}
\label{e16.4a}
      {\cal K}_{A^\ast A'} = \frac{i\tilde{\omega}_{A'}^2}
      {\hbar\varepsilon_0 c^2}\,
      {\bf d}_A^\ast \bbox{G}({\bf r}_A,{\bf r}_{A'},\tilde{\omega}_{A'}){\bf d}_{A'}
      = -{\textstyle\frac{1}{2}}\Gamma_{A^\ast A'}
      + i \delta_{A^\ast A'}.
\end{equation}
Examples of $\Gamma_{AB}$ and $\delta_{AB}$ as functions of
frequency in the vicinity of a surface-guided field resonance are
shown in Figs.~\ref{f1} and \ref{f2}, where a single-resonance
Drude-Lorentz-type dielectric has been assumed (for details,
see \cite{Ho01a}).
It is seen that near a microsphere
resonance the dipole-dipole coupling strength sensitively change
with frequency.
In Fig.~\ref{f1}, the two atoms are located
at diametrically opposite positions (outside the sphere), which
is an example of the interatomic distances being much larger
than the wavelengths.
Using Eqs.~(\ref{e8}) and (\ref{e3.7})
and approximating the microsphere resonance line by a Lorentzian,
it is not difficult to
prove that the absolute value of $\delta_{AB}$ vanishes
exactly on resonance and peaks at half widths of half
maximum. Fig.~\ref{f1}(b) reveals that for interatomic
distances that are large compared to the distances that would be
required in free space strong
dipole-dipole coupling can be realized, even for moderately
small atom-sphere distances.
For the example considered, the dipole-dipole coupling
strength achieved, e.g., at an interatomic distance of
about $20$ wavelengths can be as strong as that one in free space at
a distance of about $0.012$ wavelengths.
The relation between the strengths of dipole-dipole
and atom-field coupling sensitively depends on the
relative positions and dipole orientations
of the atoms to each other and to the microsphere.
By moving the atoms close to each other, the contribution of the
vacuum part to the Green tensor increases, which allows one
to vary $\delta_{AB}$ while keeping $\Gamma_{AB}$ almost
unchanged (see Fig.~\ref{f2}). Note that for other resonance lines,
$\Gamma_{AB}$ and $\delta_{AB}$ can switch signs \cite{Ho01a}.
Clearly, the aforegiven discussion also holds
for whispering gallery resonances.

Introducing the probability amplitudes
\begin{equation}
\label{e20}
      C_\pm (t) = 2^{-\frac{1}{2}}
      \left[ C_A(t) \pm C_B(t) \right]
      e^{\mp i\delta_{A^\ast B}t}
\end{equation}
of the superposition states
$|\pm\rangle = 2^{-1/2} \left( |U_A\rangle\pm|U_B\rangle \right)$,
from Eqs.~(\ref{e9}) we find that the equations for $C_+(t)$
and $C_-(t)$ decouple,
\begin{equation}
\label{e20.1}
        \dot{C}_{\pm}(t) =
        \int_0^t {\rm d}t' \Ints \!{\rm d}\omega \,
        K_\pm(t,t';\omega)\,e^{\mp i\delta_{A^\ast B}(t-t')}\, C_\pm(t'),
\end{equation}
\begin{equation}
\label{e20.2}
       K_\pm(t,t';\omega) = K_{A^\ast A}(t,t';\omega)
       \pm K_{A^\ast B}(t,t';\omega) .
\end{equation}
The integrodifferential equation (\ref{e20.1}) should
be solved numerically in general. Closed solutions can be
found in the limiting cases of weak and strong atom-field
interaction.

\begin{figure}[t!]
\begin{center}
\epsfxsize=.5\textwidth
\epsfbox{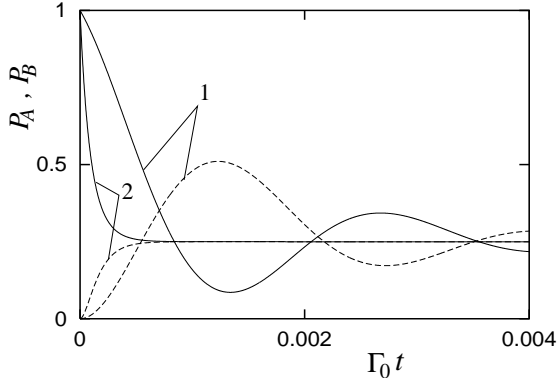}
\caption{
$P_A(t)$ (solid line) and $P_B(t)$ (dashed line)
are shown for weak resonant atom-field coupling, with the
data being taken from Fig.~\ref{f1}(a)
[curves 1: $\tilde{\omega}_A$ $\!=$ $\!1.050485\,\omega_{\rm T}$,
$\Gamma_{AA}=\Gamma_{BB}$ $\!=$ $\!640.848\,\Gamma_0$,
$\Gamma_{AB}$ $\!=$ $\!640.319\,\Gamma_0$, 
$\delta_{AB}$ $\!=$ $\!1112\,\Gamma_0$;
curves 2:
$\tilde{\omega}_A$ $\!=$ $\!1.0504867\,\omega_{\rm T}$,
$\Gamma_{AA}=\Gamma_{BB}$ $\!=$ $\!8372\,\Gamma_0$,
$\Gamma_{AB}$ $\!=$ $\!8371.5\,\Gamma_0$,
$\delta_{AB}$ $\!=$ $\!0$].
\label{f3}
}
\end{center}
\end{figure}
%
In the weak-coupling regime the Markov approximation
applies, thus
\begin{equation}
\label{e20.3}
       C_\pm(t)=2^{-1/2}e^{-\Gamma_\pm t/2},
       \qquad
       \Gamma_\pm=\Gamma_{A^\ast A}\pm \Gamma_{A^\ast B}.
\end{equation}
The upper-state occupation probabilities
\mbox{$P_{A(B)}(t)$ $\!=|C_{A(B)}(t)|^2$} then read
\begin{equation}
\label{e16.5}
         P_{A(B)}(t)= {\textstyle {1\over 2}}
         \left[ \cosh\left(\Gamma_{A^\ast B}t\right)
       +(-) \cos\left(2\delta_{A^\ast B}t\right) \right]
       e^{-\Gamma_{B^\ast B} t}.
\end{equation}
{F}rom Eq.~(\ref{e16.5}) it is seen that when
$|\delta_{A^\ast B}|$ $\!\gg$ $\!\Gamma_{B^\ast B}$, then the
excitation energy is exchanged back and forth between atoms $A$ and $B$
(Fig.~\ref{f3}, curves 1). In the opposite case of
\mbox{$|\delta_{A^\ast B}|$ $\!\ll$ $\!\Gamma_{B^\ast B}$}, $P_A$ and
$P_B$ quickly approach equal values and then decay very slowly
(curves 2). During this slow decay process, the two atoms are
partially entangled \cite{Ho01a}.
Both types of temporal evolution can also appear in free space,
but only for interatomic distances that are much smaller than
the wavelengths.

\begin{figure}[t!]
\begin{center}
\epsfxsize=\textwidth
\epsfbox{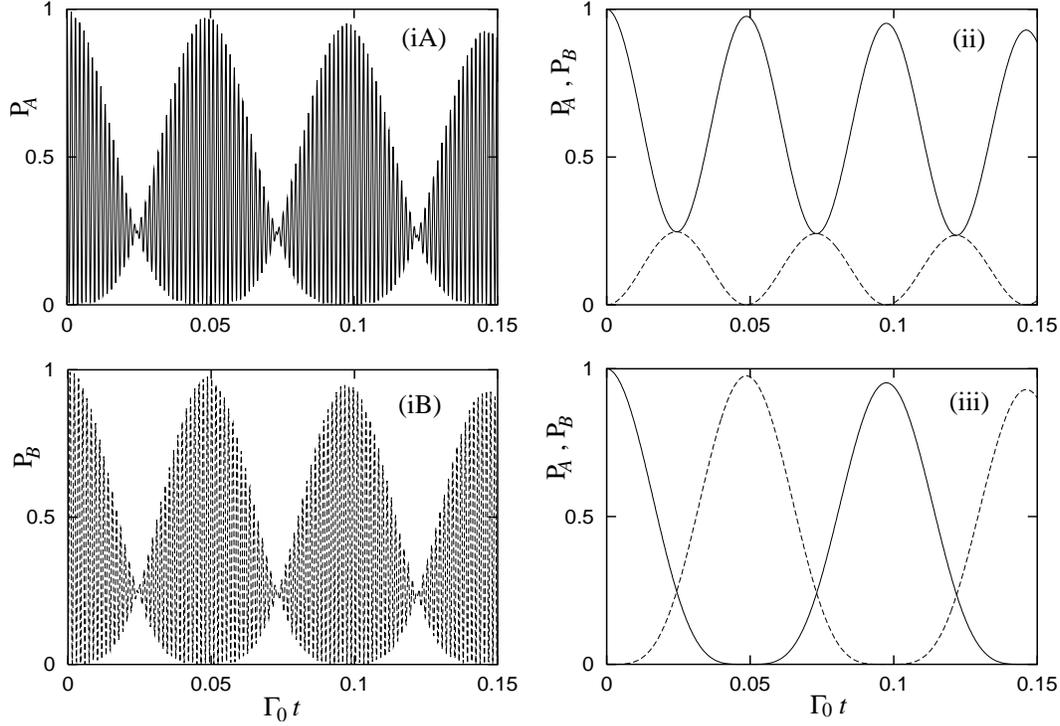}
\caption{
$P_A(t)$ (solid line) and $P_B(t)$ (dashed line)
are shown for strong resonant atom-field coupling,
with the data being taken from Figs. \ref{f1}(a) and \ref{f2}
[$\Gamma_0$ $\!=$ $\!10^{-6}\,\omega_{\rm T}$,
$\Omega_+$ $\!\simeq$ $\!128\,\Gamma_0$,
$\Delta\omega_m$ $\!\simeq$ $\!5$ $\!\times$ $\!10^{-7}\,\omega_{\rm T}$;
(i) $\tilde{\omega}_A$ $\!=$ $\!1.04835747\,\omega_{\rm T}$,
$R$ $\!\simeq$ $\!0.01\,\lambda_{\rm T}$,
$\delta_{AB}$ $\!\simeq$ $\!-$ $\!2129\,\Gamma_0$;
(ii) $\tilde{\omega}_A$ $\!=$ $\!1.05045444\,\omega_{\rm T}$,
$R$ $\!\simeq$ $\!0.027\,\lambda_{\rm T}$,
$\delta_{AB}$ $\!\simeq$ $\!-$ $\!32.2\,\Gamma_0$;
(iii) $\tilde{\omega}_A$ $\!=$ $\!\omega_m$ $\!=$
$\!1.0504867\,\omega_{\rm T}$,
$R$ $\!=$ $\!20.04\,\lambda_{\rm T}$,
$\delta_{AB}\simeq 0$].
\label{f4}
}
\end{center}
\end{figure}
%
In the strong-coupling limit, we restrict our attention to
the case when the absolute value of the two-atom term
$K_{A^\ast B}(t,t';\omega)$ is of the same order of magnitude
as the absolute value of the single-atom term
$K_{A^\ast A}(t,t';\omega)$, so that there is
a strong contrast in the magnitude of $K_+(t,t';\omega)$
and $K_-(t,t';\omega)$. As a consequence, the strong-coupling
regime can be realized for either the state $|+\rangle$ or
the state $|-\rangle$, but not for both at the same time.
Assuming that the field resonance strongly coupled to the
atoms has a Lorentzian shape, with $\omega_m$
and $\Delta\omega_m$ being the central frequency and
the half width at half maximum respectively, we can
perform the frequency integral in Eq.~(\ref{e20.1}) in
a closed form, on extending it to $\pm\infty$. The 
further calculation can then be performed as described
in \cite{Ho01a}. In particular for exact resonance, i.e.,
\mbox{$\omega_m$ $\!=$ $\!\tilde{\omega}_A\mp\delta_{A^\ast B}$},
we derive
\begin{equation}
\label{e21}
       C_\pm(t) = 2^{-\frac{1}{2}}
       e^{-\Delta\omega_m t/2} \cos\!\left(\Omega_\pm t/2\right),
      \qquad
      \Omega_\pm = \sqrt{2\Gamma_\pm\Delta\omega_m}\,.
\end{equation}
For the probability amplitudes of the remaining states
$|\mp\rangle$, which are weakly coupled to the field, we have
$C_\mp(t)$ $\!=$ $\!2^{-1/2} e^{-\Gamma_\mp t/2}$.
It then follows that
\begin{equation}
       P_{A(B)}(t)=\textstyle{1\over 4} \Bigl[
       e^{-\Gamma_\mp t} + e^{-\Delta\omega_m t}
       \cos^2\!\left(\Omega_\pm t/2\right)
\label{e22.1}
     +(-) \,2e^{-(\Delta\omega_m  +\Gamma_\mp)t/2}
     \cos\!\left(\Omega_\pm t/2\right)
     \cos\!\left(2\delta_{A^\ast B}t\right) \Bigr].
\end{equation}
The upper (lower) signs refer to the case where the state
$|+\rangle$ ($|-\rangle$) is strongly coupled to the medium-assisted
field. Typical examples are shown in Fig.~\ref{f4}. When the
atoms are sufficiently close to each other
($4|\delta_{A^\ast B}|$ $\!\gg$ $\!\Omega_\pm$),
then a beating-type behavior of $P_A(t)$ and $P_B(t)$ as shown
in Figs.~\ref{f4}(iA) and (iB), respectively, may be observed.
The beat is between the oscillation of the frequency
$2\delta_{A^\ast B}$, which arises from the two-atom dipole-dipole
coupling, and the Rabi oscillation of frequency $\Omega_\pm$.
Recall that
the collapses and revivals of the atomic level populations
in the Jaynes-Cummings model are caused by the
presence of oscillations with noncommensurate single-atom
Rabi frequencies.
Partial trapping of the excitation energy
in atom $A$ as shown  in Fig.~\ref{f4}(ii) may be observed if
the distance between the atoms is slightly increased ($4|\delta_{A^\ast B}|$
$\!\simeq$ $\!\Omega_\pm$). This trapping can be understood as
resulting from a destructive
interference between the two channels of energy transfer, one via
virtual and the other via real medium-assisted field excitation.
Finally, Fig.~\ref{f4}(iii) shows that for larger
distances the motion becomes governed by the Rabi oscillations
($4|\delta_{A^\ast B}|$ $\!\simeq$ $\!0$ $\!\ll$ $\!\Omega_\pm$).

\section{Outlook}
\label{sec5}

The present analysis has left a number of open questions,
on which future work will concentrate. Depending upon
the strengths of the resonant dipole-dipole
coupling and the resonant atom-field coupling and
their relations to each other, multiplet spectra of the
emitted light can be expected. Further, the
problem of mutual interaction of atoms whose transition
frequencies must be regarded as being different with regard to the
variation of the medium-assisted Green tensor needs
special emphasis, because in such a case the effective
Hamiltonian (\ref{e11}) does not apply.

\vspace{2ex}
\noindent
{\bf Acknowledgement:}
This work was supported by the Deutsche Forschungsgemeinschaft.


\end{document}